\documentclass[11pt,a4paper]{article}

\usepackage[margin=1in]{geometry}
\usepackage{caption}
\usepackage{graphicx}
\usepackage{tabularx}
\usepackage[normalem]{ulem}
\usepackage{makecell}
\usepackage{xspace}
\usepackage{booktabs}
\usepackage{multirow}
\usepackage{lipsum}
\usepackage{enumitem}
\usepackage{amsmath}
\usepackage{float}
\usepackage{adjustbox}
\setlist[itemize]{noitemsep, topsep=0pt}
\usepackage{subcaption}
\usepackage{algorithmicx}
\usepackage{algorithm}     % 알고리즘 틀
\usepackage{algpseudocode} % 알고리즘 구문
\usepackage{booktabs}   
\usepackage{pifont}     % 체크박스 심볼(\ding)
\usepackage{xcolor}     
\usepackage{colortbl}   

\usepackage{amssymb}
\usepackage{soul}
\usepackage{hyperref}
\usepackage{cleveref}

\hypersetup{
    colorlinks=true,
    linkcolor=blue,
    filecolor=magenta,
    urlcolor=cyan,
    citecolor=blue
}

\definecolor{Gray}{gray}{0.9}

\begin{document}

\title{\textbf{Proactive Computing}}

\author{
    \textbf{Joonhee Lee}\\
    Seoul, South Korea\\
    \texttt{neo81389@yonsei.ac.kr}
}

\date{} % 날짜를 비워두거나 \date{\today} 사용

\maketitle

\begin{abstract}
    Computing systems are moving from reactive tools toward systems that sense, interpret, predict, and act before explicit user requests. This transition is enabled by the global scale of mobile connectivity, the rapid expansion of wearable and ambient sensing, advances in machine learning and foundation models, distributed edge infrastructure, and physical actuation. We define \emph{proactive computing} as a paradigm in which systems infer user context, anticipate future needs or risks, and initiate information delivery or actions at an appropriate time. This survey distinguishes proactive computing from reactive, context-aware, adaptive, and predictive computing, and frames proactivity as a system-level integration problem across sensing, understanding, decision making, action, and governance. We review the technological enablers of proactive computing, organize its design space, analyze technical challenges such as uncertainty-aware triggering and the prediction-to-action gap, and discuss socio-technical issues involving user acceptance, trust, privacy, accountability, fairness, and sustainability. We argue that the key research challenge is not merely improving prediction accuracy, but determining when, how, and whether systems should act on behalf of users.
\end{abstract}

\section{Introduction}
\label{sec:Intro}

Computing has progressively moved closer to human life. 
Early computing systems were largely stationary and reactive, requiring users to explicitly request information or initiate computation. 
The vision of ubiquitous computing argued that computation would eventually disappear into everyday environments rather than remain confined to desktop machines \cite{weiser1991computer}. 
The emergence of mobile computing changed this paradigm by enabling sensing, communication, and processing on portable devices such as smartphones, tablets, and embedded devices. 
This transition is no longer only conceptual: Ericsson reports approximately 8.5 billion mobile subscriptions and 7.4 billion smartphone subscriptions in 2025, with 5G subscriptions reaching about 2.9 billion by the end of that year \cite{ericsson2026mobility}. 
A large body of research has since investigated how small, resource-constrained, and mobile devices can sense the physical world, process data locally, and support context-aware services \cite{dey2001understanding}.

However, recent technological trends suggest that the conventional mobile computing paradigm is no longer sufficient. 
First, sensing devices have become increasingly personal and pervasive. 
Beyond smartphones, wearable devices such as smartwatches, smart rings, earbuds, and health sensors continuously capture fine-grained information about users' activities, physiological states, locations, and surrounding environments. 
IDC (International Data Corporation) estimates that global wearable device shipments reached 611.5 million units in 2025, growing 9.1\% year over year, and forecasts 689.7 million units by 2030 \cite{idc2026wearables}. 
Second, advances in machine learning and foundation models have substantially improved the ability of computing systems to infer high-level context from noisy and heterogeneous sensor data, while also introducing new concerns such as hallucination, bias, and unreliable reasoning in multimodal models \cite{bommasani2021foundation,bai2024hallucination}. 
Third, distributed computing infrastructures, including edge computing, 6G networks, and Open RAN, are expanding computation beyond individual devices by allowing mobile systems to collaborate with nearby infrastructure \cite{satyanarayanan2017edge,mach2017mobile}. 
Finally, advances in robotics and actuation technologies are enabling computing systems not only to provide information but also to physically assist users in the real world.

These trends point toward a new computing paradigm: \emph{proactive computing}. 
The term has roots in Tennenhouse's call for systems that ``anticipate our needs and take action on our behalf'' rather than simply wait for commands \cite{tennenhouse2000proactive}. 
Unlike reactive systems that respond only after explicit user requests, proactive computing systems infer user context, anticipate future needs, and initiate information delivery or actions before users explicitly ask. 
Examples include a wearable system that warns a user before health risk increases, a mobile assistant that prepares relevant information before a meeting, a smart home system that adjusts the environment based on predicted comfort, or a robot that assists a user before a task becomes difficult.

Despite its promise, proactive computing remains difficult to define and systematize. 
Many existing systems can sense context or predict future states, but prediction alone does not constitute proactivity. 
A proactive system must additionally decide when to act, how to act, and whether acting is appropriate. 
This introduces new technical and socio-technical challenges involving personalization, uncertainty, user acceptance, trust, privacy, accountability, and sustainability.

This paper reviews proactive computing as an emerging socio-technical computing paradigm. 
Rather than treating proactive computing as a single algorithmic problem, we frame it as a system-level integration problem across sensing, intelligence, infrastructure, and action. 
We first discuss the conceptual evolution from reactive computing to context-aware, predictive, and proactive computing. 
We then summarize the technological enablers of proactive computing, present a design space for proactive systems, analyze key technical and socio-technical challenges, and outline future research directions.

The contributions of this survey are as follows:
\begin{itemize}
    \item We define proactive computing as a computing paradigm that integrates sensing, understanding, prediction, and action before explicit user requests.
    \item We distinguish proactive computing from reactive, context-aware, adaptive, and predictive computing.
    \item We organize the design space of proactive computing around sensing, intelligence, personalization, autonomy, action, deployment, and governance.
    \item We identify key technical challenges, including the gap between low-level sensing and high-level understanding, the prediction-to-action gap, uncertainty-aware decision making, and resource constraints.
    \item We discuss socio-technical issues such as user acceptance, trust, privacy, accountability, cultural differences, and sustainability.
\end{itemize}

\section{From Reactive to Proactive Computing}
\label{sec:evolution}

\subsection{Reactive Computing}

Traditional computing systems are primarily reactive. 
In a reactive paradigm, users explicitly formulate requests, and systems respond to those requests. 
Examples include search engines, command-line interfaces, graphical user interfaces, and conventional mobile applications. 
The system remains passive until the user initiates interaction.

This paradigm gives users direct control, but it also places the burden of initiation on them. 
Users must recognize their own needs, formulate appropriate queries, and decide when to interact with the system. 
As computing becomes embedded into daily life, this model becomes increasingly limited. 
Many useful services require timely intervention before users explicitly recognize or express their needs.

\subsection{Context-Aware Computing}

Context-aware computing extends reactive computing by allowing systems to sense and adapt to the user's current situation. Dey's widely used definition characterizes context as any information that can describe the situation of an entity, and context-aware systems as systems that use such information to provide relevant services \cite{dey2001understanding}. 
A context-aware system may use information such as location, activity, time, device state, or environmental conditions to modify its behavior. 
For example, a smartphone may silence notifications during a meeting, or a navigation application may adjust routes based on current location.

Context-aware computing represents an important step toward more intelligent interaction. 
However, many context-aware systems remain reactive or rule-based. 
They typically recognize the current context and adapt predefined behaviors, but they do not necessarily anticipate future needs or initiate meaningful actions.

\subsection{Predictive and Anticipatory Computing}

Predictive computing goes beyond current context recognition by estimating future states. Anticipatory mobile computing formalizes this idea by combining mobile sensing and machine learning to reason about future events and to support decision making before the event occurs \cite{pejovic2015anticipatory}. 
For example, a system may predict a user's next location, future activity, health risk, mobility pattern, or information need. 
Such prediction enables anticipatory services that prepare resources or recommendations before they are explicitly requested.

However, prediction alone is not equivalent to proactivity. 
A system that predicts a user's future state but does not decide whether or how to act remains incomplete from the perspective of proactive computing. 
The key challenge is not only to predict what will happen, but also to determine what the system should do in response.

\subsection{Proactive Computing}

We define proactive computing as a paradigm in which computing systems infer user context, anticipate future needs or risks, and initiate information delivery or actions without explicit user requests. This definition extends the original proactive computing vision by requiring an explicit connection between anticipation and situated action \cite{tennenhouse2000proactive,pejovic2015anticipatory}. 
This definition emphasizes three elements: sensing, understanding, and action. 
A proactive system must observe relevant signals, interpret them in relation to user needs, and intervene at an appropriate time.

The distinction between related paradigms can be summarized as follows:
\begin{table}[t]
\centering
\caption{Conceptual progression from reactive to proactive computing. The key boundary is the transition from recognizing or predicting context to deciding whether an intervention should occur.}
\label{tab:reactive_proactive_progression}
\begin{tabular}{p{0.22\linewidth}p{0.30\linewidth}p{0.34\linewidth}}
\toprule
Paradigm & Main capability & Typical limitation \\
\midrule
Reactive computing & Responds after explicit user input & Requires the user to notice a need and initiate interaction \\
Context-aware computing & Adapts to the current situation using contextual information & Often remains rule-based and focused on present context \\
Predictive computing & Estimates future states, risks, or needs & Does not necessarily decide whether action is appropriate \\
Proactive computing & Decides when and how to act before explicit user requests & Must balance usefulness, timing, uncertainty, user control, and safety \\
\bottomrule
\end{tabular}
\end{table}

Thus, the central challenge of proactive computing is the transition from prediction to action. 
A proactive system must determine not only what is likely to happen, but also whether an intervention is useful, acceptable, trustworthy, and safe.

\section{Technological Enablers of Proactive Computing}
\label{sec:enablers}

\subsection{Personal and Pervasive Sensing}

The first major enabler of proactive computing is the evolution of sensing technologies. Modern users carry and wear multiple sensing devices, including smartphones, smartwatches, smart rings, earbuds, cameras, health sensors, ambient IoT devices, and other embedded devices. These devices continuously capture signals related to movement, location, physiology, social interaction, environmental conditions, and device usage. The scale of this sensor substrate is substantial: global wearable shipments reached 611.5 million units in 2025, and IDC projects approximately 689.7 million units by 2030 \cite{idc2026wearables}. At the ambient scale, IoT Analytics estimates that connected IoT devices will grow from 18.5 billion in 2024 to 21.1 billion in 2025 and 39 billion in 2030, while Wi-Fi, Bluetooth, and cellular links account for 32\%, 24\%, and 22\% of 2025 IoT connections, respectively \cite{iotanalytics2025connections}. This growth means that proactive systems can be built on increasingly continuous, multimodal, and personally situated data streams rather than on occasional user input alone.

Within this pervasive sensing layer, brain–computer interfaces (BCIs) and neurotechnology can also be integrated as an emerging modality of multi-channel, high-resolution time-series data. Unlike inertial or environmental sensors that infer user states indirectly, non-invasive or embedded neural interfaces capture cognitive workload, attention, and immediate motor intentions before they manifest as overt behavioral actions \cite{willett2021handwriting,willett2023speech}. Although still transitioning from clinical and specialized environments to broader applications, treating these neural signals as a high-bandwidth temporal sensing channel allows proactive systems to capture subtle indicators of user affect or intent that are otherwise invisible to external hardware.

This shift enables computing systems to obtain more fine-grained and personal information than traditional desktop or smartphone-only systems. Wearable and physiological sensors capture continuous behavioral rhythms, ambient sensors track environmental context, and neural interfaces monitor immediate cognitive correlates. Together, these multimodal sensing streams provide the foundational data fabric for inferring complex user states and situational needs.

However, sensing alone is insufficient. Low-level sensor signals must be converted into high-level semantic understanding. For proactive computing, recognizing that a user is walking or sitting may not be enough; the system may need to infer whether the user is stressed, distracted, at risk, available, intending to act, or in need of assistance.

\begin{figure*}[t]
\centering
\includegraphics[width=\textwidth]{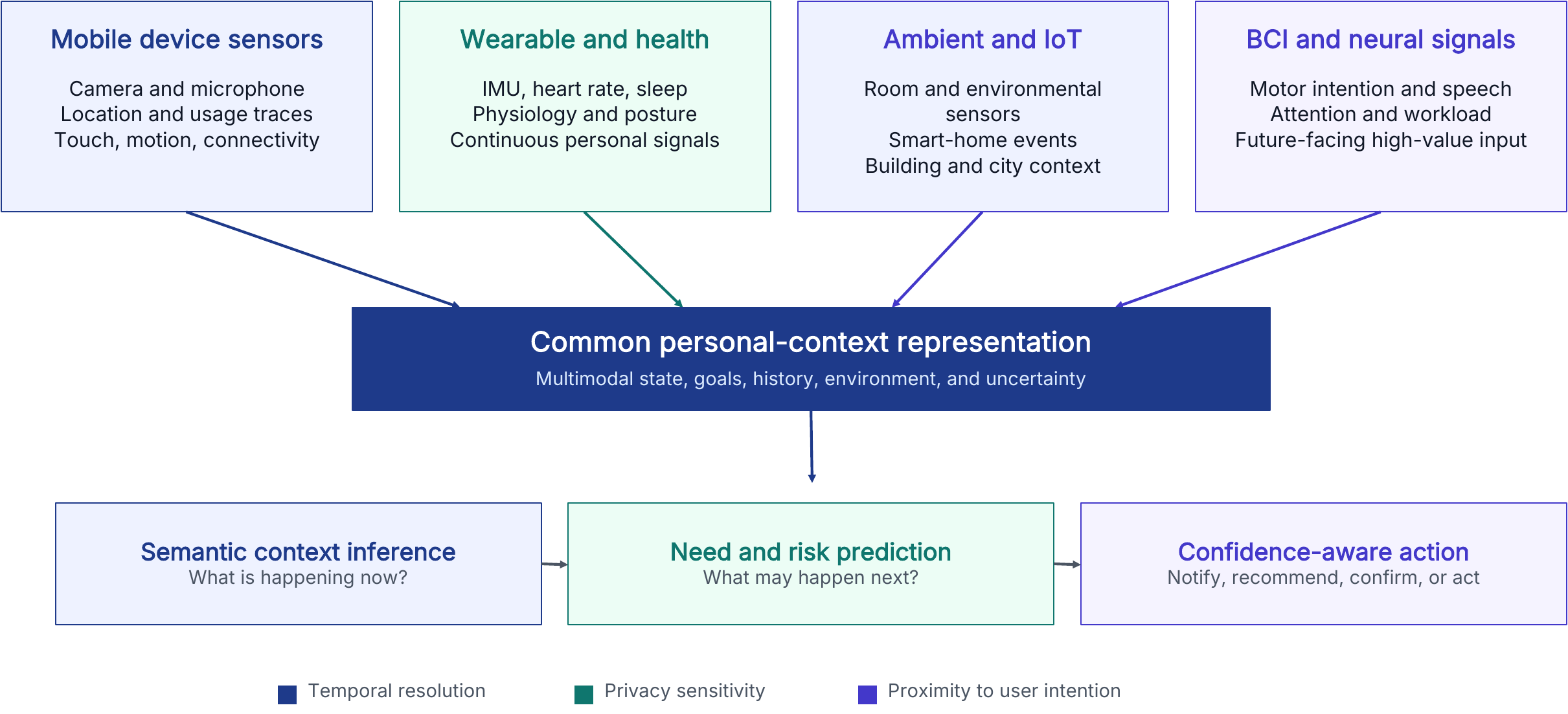}
\caption{Multimodal sensing stack for proactive computing. The figure shows the sensing layers---mobile, wearable, ambient/IoT, and neural---and illustrates that each layer contributes different temporal resolution, privacy sensitivity, and semantic proximity to user intention.}
\label{fig:sensing_stack}
\end{figure*}

Figure~\ref{fig:sensing_stack} visually connects smartphone, wearable, ambient, and neural sensing to proactive inference.

\subsection{AI-Based Context Understanding}

The second enabler is the advancement of AI-based understanding. Machine learning has improved the ability of systems to infer activities, intentions, preferences, and future states from heterogeneous sensor data. Recent advances in representation learning, multimodal learning, self-supervised learning, and foundation models further expand the ability of systems to reason over complex context \cite{bommasani2021foundation}. The scale of this transition is visible in widely used foundation-model milestones: GPT-3 demonstrated 175-billion-parameter language modeling for few-shot adaptation, and CLIP trained vision-language representations on 400 million image–text pairs, enabling zero-shot transfer across more than 30 computer-vision datasets \cite{brown2020language,radford2021learning}. At the same time, deep learning has become a major research direction for wearable human activity recognition, where surveys identify robustness across users, devices, and domains as a central barrier for real-world deployment \cite{zhang2022har}. Foundation models do not remove the need for system-level safeguards: hallucination remains a documented limitation in multimodal large language models, especially when systems must ground outputs in visual or sensor evidence \cite{bai2024hallucination}.

In proactive computing, AI serves several roles. First, it transforms raw sensor data into semantic representations. Second, it predicts future states or needs. Third, it supports decision making by estimating the usefulness, risk, and timing of possible actions. Finally, it enables personalization by adapting models to individual users.

Nevertheless, AI-based understanding introduces uncertainty. Sensor data are noisy, user behavior is diverse, and future needs are difficult to predict. Therefore, proactive systems require not only accurate models but also mechanisms for uncertainty estimation, confidence-aware triggering, and graceful failure.

\subsection{Distributed and Edge Infrastructure}
\label{subsec:distributed_edge}

The third enabler is distributed computing infrastructure. Mobile and wearable devices remain constrained by battery, memory, computation, and thermal limits. At the same time, proactive services often require continuous sensing, real-time inference, personalization, and sometimes large AI models. The relevant design question is therefore not simply whether inference should be local or remote. It is how a proactive system should continuously place sensing, feature extraction, semantic inference, memory retrieval, policy selection, and model update across a hierarchy of compute locations.

The first layer is \emph{device-only} execution. It is the most private and robust to network failures, and it is appropriate for wake-word detection, anomaly detection, compact HAR, safety monitors, and low-risk interaction gating. The second layer is a \emph{personal hub}, usually a smartphone or robot edge computer, that aggregates nearby wearables and ambient sensors. The third layer is \emph{on-premise edge} infrastructure in homes, hospitals, factories, campuses, vehicles, or buildings. This layer is important for proactive computing because it can keep sensitive data inside an administrative boundary while offering more memory, power, and cooling than a watch or phone. The fourth layer is \emph{network-edge offloading}, including cloudlets, multi-access edge computing (MEC), and base-station AI. Satyanarayanan argues that cloudlet-style edge computing is essential when tight latency control is required and notes that such systems target end-to-end latencies on the order of a few tens of milliseconds \cite{satyanarayanan2017edge}. Kang et al.'s Neurosurgeon system similarly reports a 3.1\(\times\) average latency improvement when DNN computation is partitioned between mobile devices and cloud or edge resources \cite{kang2017}. At the standards level, ETSI positions MEC as a way to bring application hosting from centralized data centers to the network edge and to expose local context and real-time network information through standardized APIs \cite{etsi2018mec5g}.

The most recent industrial version of this network-edge idea is base-station AI or AI-RAN. In this model, the radio access network is not only a communication pipe but also a nearby AI execution substrate. NVIDIA reports that a SoftBank outdoor 5G AI-RAN field trial in Fujisawa City integrated 20 radio units, a 5G core, and 100 mobile user equipments on GH200-based infrastructure; each GH200 server processed 20 5G cells with 100-MHz bandwidth in RAN-only mode, and outdoor tests demonstrated 816~Mbps per cell with carrier-grade availability \cite{nvidia2024airan}. The same report states that AI and RAN multi-tenancy can target nearly 100\% utilization compared with roughly 33\% typical RAN-only utilization, and illustrates local AI-RAN inference for robotics and factory multimodal AI \cite{nvidia2024airan}. These are vendor-reported results rather than peer-reviewed general laws, but they show that telco infrastructure is being reimagined as a deployable AI substrate for latency-sensitive proactive services.

The final layer is regional or public cloud. Cloud resources remain necessary for large VLM/LLM reasoning, cross-user analytics, foundation-model training, and long-context retrieval, but they are less suitable for highly private, always-on, or safety-critical sensing loops. Thus, a proactive system should treat cloud, on-premise edge, MEC/base-station AI, phone hubs, and wearables as a staged continuum. The architectural objective is an \emph{edge continuum}: not a set of disconnected computers, but a single logical computing fabric in which data locality, privacy, latency, energy, model size, and risk determine where each inference step executes. In such a fabric, repeaters, base stations, home gateways, local servers, smartphones, and edge devices act as parts of one computer from the user's perspective.

\begin{figure*}[t]
\centering
\includegraphics[width=0.95\textwidth]{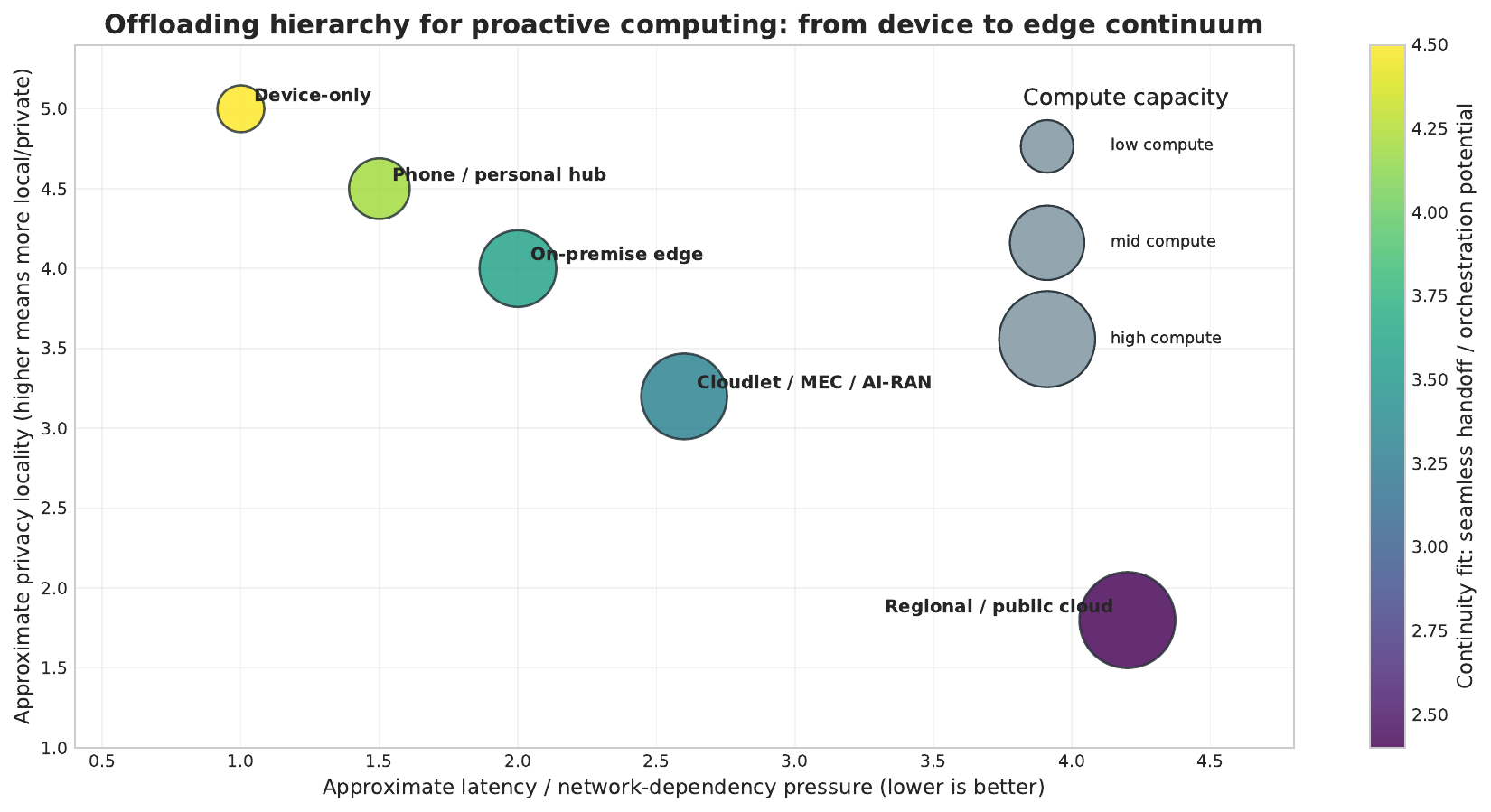}
\caption{Offloading hierarchy for proactive computing. Bubble size indicates approximate compute capacity, the horizontal axis indicates latency and network-dependency pressure, and the vertical axis indicates privacy locality. The plotted scores are a conceptual synthesis for design comparison rather than a new benchmark; quantitative anchors include cloudlet tens-of-milliseconds latency targets, Neurosurgeon's 3.1\(\times\) latency improvement through DNN partitioning, ETSI MEC's standardized network-edge context exposure, and vendor-reported AI-RAN field-trial evidence \cite{satyanarayanan2017edge,kang2017,etsi2018mec5g,nvidia2024airan}.}
\label{fig:offloading_hierarchy_tradeoff}
\end{figure*}

This continuum also creates limits and open directions. Offloading can leak sensitive context, fail under mobility or congestion, increase tail latency, and blur accountability when an action depends on several administrative domains. Future proactive systems therefore need placement policies that are uncertainty-aware and privacy-aware, cryptographic or federated personalization when raw data cannot move, standardized APIs for local context exposure, and fallback behavior that keeps safety-critical functions local when the network or remote model is unavailable. Figure~\ref{fig:offloading_hierarchy_tradeoff} summarizes the intended hierarchy and the central trade-off: compute capacity generally increases as one moves outward, while privacy locality and deterministic latency generally decrease.

\subsection{Human-Activity, Time-Series, and Physical Context Understanding}
\label{subsec:har_timeseries_physical}

The fourth enabler is the improvement of context-understanding models for signals that are central to proactive computing: human activity, multimodal vision, physiological and behavioral time series, and physical-world understanding. Vision-language encoders are comparatively mature. CLIP trained on 400 million image–text pairs and demonstrated zero-shot transfer across more than 30 computer-vision datasets \cite{radford2021learning}. However, proactive computing cannot rely on image encoders alone. It needs models that understand wearable motion, physiological rhythms, sparse smart-home events, long-horizon routines, and physical interaction with objects.

Human activity recognition (HAR) illustrates both progress and incompleteness. Classical wearable HAR has moved from hand-crafted features to deep learning and representation learning, but surveys still identify cross-user, cross-device, and cross-domain generalization as central barriers \cite{zhang2022har}. Recent foundation-model work is changing the scale of the field. Yuan et al. trained self-supervised HAR representations using 700,000 person-days of wearable data, showing that population-scale unlabeled sensing can improve downstream activity recognition \cite{yuan2024selfsupervisedhar}. Sensor-oriented foundation-model surveys now describe UniMTS, SensorLM, sensor-to-text models, RF/IMU-to-vision-language alignment, masked signal modeling, and language-grounded sensing as an emerging HAR foundation-model ecosystem \cite{haresamudram2026fmhar,zhang2024unimts,zhang2025sensorlm}. Yet the same surveys emphasize that sensor data remain fragmented across device types, sampling rates, body placements, label taxonomies, and privacy regimes, which means that IID accuracy on a single dataset is insufficient for proactive deployment \cite{haresamudram2026fmhar}.

Time-series foundation models show a similar pattern. A recent IEEE TPAMI survey organizes self-supervised time-series learning into contrastive, generative, predictive, and hybrid paradigms and concludes that robust transfer across domains remains a key research problem \cite{zhang2024sslts}. For proactive computing, this matters because many useful inferences are not instantaneous labels but temporal judgments: whether a user is becoming fatigued, whether a routine is deviating, whether an intervention should be delayed, or whether an environmental pattern predicts future discomfort. These tasks require long-horizon temporal abstraction, calibrated uncertainty, and personalization rather than only short-window classification.

The arXiv position paper \emph{AI Should Sense Better, Not Just Scale Bigger} provides a useful bridge between these observations and proactive computing. It argues that AI systems should adapt sensing at the input level, not only scale models and datasets, and reports that the Lens adaptive-sensing prototype improves accuracy by up to 47.58 percentage points without modifying the model; it also notes that a 5M-parameter EfficientNet-B0 can surpass a 632M-parameter OpenCLIP-H trained on 160\(\times\) more data under ideal sensor adaptation \cite{baek2025adaptive,baek2025adaptive_sensing_position}. This view is highly aligned with proactive computing. If context understanding is partly limited by how the world is sensed, then proactive agents should not merely run larger models on fixed sensor streams. They should adapt sampling rate, exposure, gain, modality selection, sensor placement, compression, and offloading decisions before the signal becomes irreversibly lossy.

Physical understanding is the least mature but most safety-critical part of this stack. PhysBench evaluates VLM physical-world understanding with 10,002 interleaved video-image-text entries across four major domains, 19 subclasses, and 8 capability dimensions; experiments on 75 representative VLMs find that models still struggle with physical-world understanding despite stronger common-sense reasoning \cite{chow2025physbench}. This gap is important because proactive systems that control robots, vehicles, smart homes, or assistive devices must anticipate physical consequences, not only recognize semantic labels. A wrong physical prediction can turn a proactive action into a safety hazard.

\begin{figure*}[t]
\centering
\includegraphics[width=0.94\textwidth]{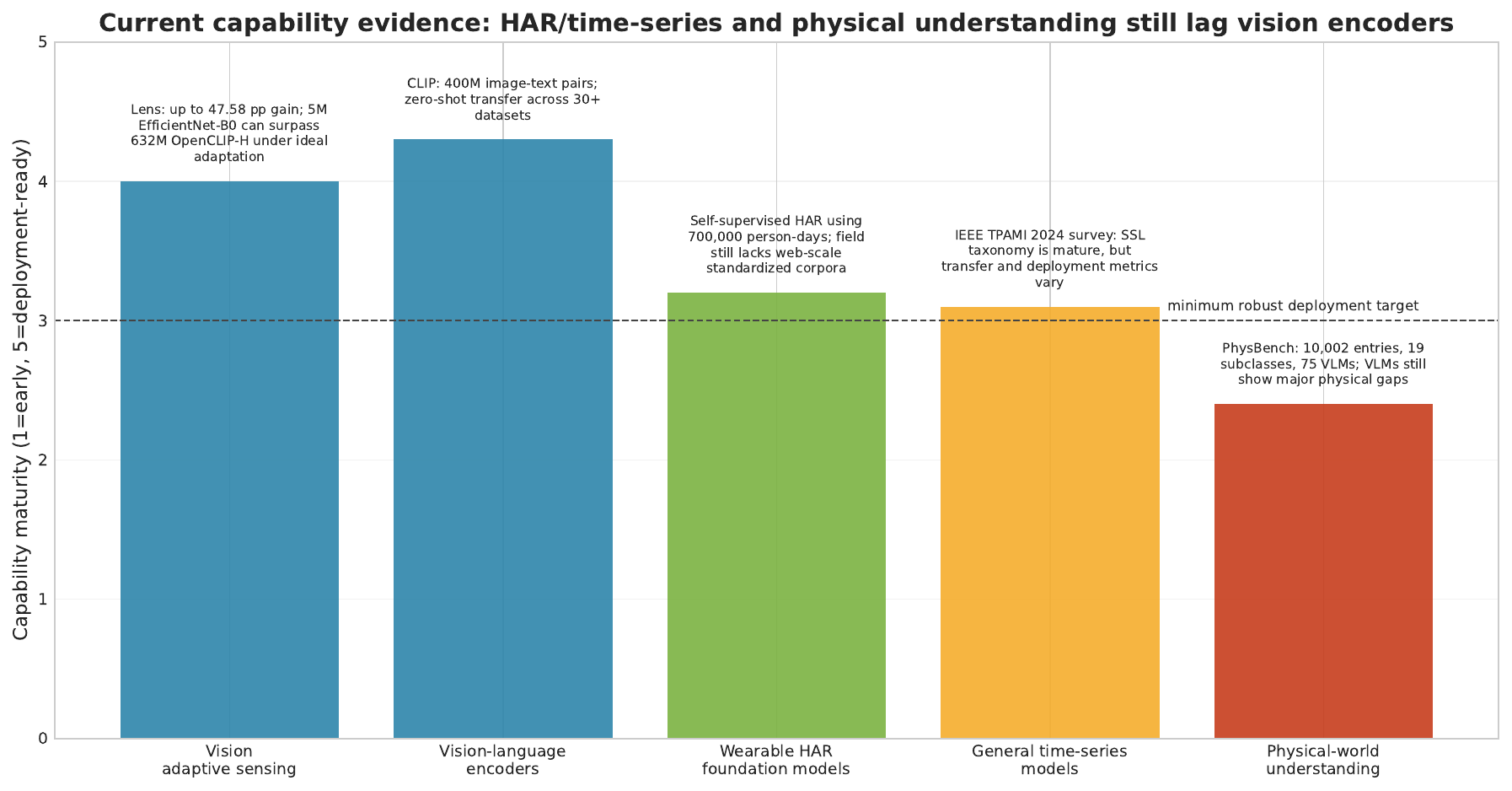}
\caption{Capability evidence for proactive context understanding. The maturity scores are an interpretive synthesis for this survey rather than benchmark results. The labels report quantitative anchors: adaptive sensing can improve accuracy by up to 47.58 percentage points and allow a 5M-parameter model to compete with a 632M-parameter encoder under ideal adaptation; CLIP used 400 million image–text pairs; wearable HAR has reached 700,000 person-days of self-supervised data; time-series self-supervision has a mature taxonomy but unresolved transfer and deployment metrics; and PhysBench evaluates 75 VLMs on 10,002 physical-understanding entries \cite{baek2025adaptive,baek2025adaptive_sensing_position,radford2021learning,yuan2024selfsupervisedhar,zhang2024sslts,chow2025physbench}.}
\label{fig:context_understanding_evidence}
\end{figure*}

Figure~\ref{fig:context_understanding_evidence} summarizes the state of evidence. The conclusion is deliberately conservative: vision encoders and adaptive sensing already provide strong building blocks, HAR and time-series foundation models are promising but still fragmented, and physical-world understanding requires substantial improvement before high-stakes proactive actuation can be trusted.

\subsection{Device Capability and On-Device Model Fit}
\label{subsec:device_model_fit}

The feasibility of proactive computing also depends on whether the sensing device, wearable, or robot can run the required model under realistic resource constraints. Current hardware already supports limited local intelligence, but the deployment envelope differs sharply across device classes. Flagship smartphones represent the strongest consumer mobile baseline: the Samsung Galaxy S25 Ultra provides 12~GB RAM and storage configurations from 256~GB to 1~TB, while Qualcomm reports that the Snapdragon 8 Elite improves AI performance and AI performance per watt by 45\% relative to the previous generation \cite{samsung2025galaxys25,qualcomm2024snapdragon8elite}. Apple's iPhone 16 Pro uses the A18 Pro with a 16-core Neural Engine, 128~GB--1~TB storage, a 17\% increase in total system memory bandwidth, and Apple Intelligence support through on-device processing with Private Cloud Compute when needed \cite{apple2024iphone16pro}. Wearables and glasses are more constrained: Apple Watch Series~10 includes an S10 SiP with a 64-bit dual-core processor, a 4-core Neural Engine, 64~GB capacity, and up to 18~hours of normal battery life, whereas Ray-Ban Meta smart glasses provide a Snapdragon AR1 Gen~1 platform, 12~MP camera, 1080p video up to 60~seconds, and a five-microphone array, but neither vendor discloses RAM or TOPS for these wearable classes \cite{apple2024watchseries10,meta2023rayban}. Robot and embedded edge platforms offer a much larger local inference envelope: NVIDIA's Jetson Orin Nano Super provides 8~GB memory, 67~INT8 TOPS, 102~GB/s memory bandwidth, and a 7--25~W power range; the larger Jetson Orin family reaches up to 275~TOPS for autonomous machines and robotics \cite{nvidia2024orinnanosuper,nvidia2024jetsonorin}.

Table~\ref{tab:hardware_capability_matrix} summarizes these device classes as a capability matrix. The key interpretation is that smartphones can plausibly host small quantized LLMs, local ASR, lightweight image encoders, and routing policies, but continuous proactive agents must share memory, compute, battery, and thermal headroom with the operating system and foreground user applications. Watches and glasses are better treated as always-available sensing and interaction endpoints whose local AI should focus on wake words, anomaly detection, sensor fusion, compact classifiers, and embedding extraction, while heavier reasoning should be delegated to the phone, edge, or cloud. Robots and embedded edge boxes can run richer perception and planning loops locally, but their higher power budget does not remove the need for safety monitors and fallback policies.

\begin{table*}[t]
\centering
\caption{Hardware capability matrix for proactive mobile, wearable, and robotic AI. The table compares representative smartphones, watches, smart glasses, and robot edge platforms by memory/storage, AI compute indicators, and plausible local proactive-AI fit. The table should be read conservatively because some vendors do not disclose RAM or TOPS, and installed memory is not equivalent to free runtime memory.}
\label{tab:hardware_capability_matrix}
\footnotesize
\begin{tabularx}{\textwidth}{>{\columncolor{gray!20}\bfseries}p{0.14\textwidth} p{0.18\textwidth} X X X}
\toprule
\rowcolor{gray!50}
Device class & Representative hardware & Memory / storage & AI compute indicator & Local proactive AI fit \\
\midrule
Flagship smartphone & Samsung Galaxy S25 Ultra / Snapdragon 8 Elite & 12 GB RAM; 256 GB--1 TB storage & Hexagon NPU: +45\% AI perf., +45\% perf/W vs prior gen. & Small LLMs (1B--4B), on-device ASR, image encoders, lightweight VLM routing; larger VLMs need offload \\
\addlinespace
Smartphone ecosystem baseline & iPhone 16 Pro / A18 Pro & 128 GB--1 TB storage; RAM not disclosed & 16-core Neural Engine; A18 Pro: +15\% CPU, +20\% GPU vs A17 Pro & Local multimodal feature extraction and small language tasks; cloud/edge handoff for heavy reasoning \\
\addlinespace
Smartwatch & Apple Watch Series 10 & 64 GB storage; RAM not disclosed & S10 SiP: 64-bit dual-core + 4-core Neural Engine & Sensor fusion, anomaly detection, wake-word/on-device Siri; LLM/VLM via phone/edge \\
\addlinespace
Smart glasses & Ray-Ban Meta smart glasses & RAM/TOPS not disclosed & Snapdragon AR1 Gen 1; 12 MP camera; 1080p video; 5 microphones & Always-available capture and voice interaction; visual-language reasoning via phone/cloud/edge \\
\addlinespace
Mobile robot / edge box & NVIDIA Jetson Orin Nano Super Dev. Kit & 8 GB memory & Up to 67 INT8 TOPS; 102 GB/s memory bandwidth & Real-time perception, small VLM/LLM planning loops, multimodal fusion, local fallback when offline \\
\addlinespace
Robot / autonomous machine & NVIDIA Jetson AGX Orin & Up to 64 GB memory & Up to 275 TOPS & Multi-camera perception, embodied planning, VLM-assisted navigation, and local safety monitors \\
\bottomrule
\end{tabularx}
\end{table*}

Model-side trends are moving in the opposite direction: models are becoming smaller, more selectively activated, and more deployable through optimization. Meta reports that Llama~3.2 includes 1B and 3B text models designed for selected edge and mobile devices, with 128K-token context, Arm optimization, and Qualcomm/MediaTek support; the lightweight variants were produced through structured pruning and knowledge distillation from larger Llama models, followed by supervised fine-tuning, rejection sampling, and DPO \cite{meta2024llama32,meta2024llama32card}. Google describes Gemma~3n as a multimodal model family optimized for phones, laptops, and tablets; it combines MobileNet-V5 vision encoding, Per-Layer Embedding caching, MatFormer nested activation, and conditional parameter loading, reducing E2B execution from more than 5B loaded parameters to an effective memory load of 1.91B parameters \cite{google2025gemma3n}. These examples illustrate why proactive computing should not be framed as either fully local or fully cloud-based. Instead, practical systems will use quantization, pruning, distillation, caching, early exiting, conditional computation, retrieval, and split execution to keep time-sensitive perception and low-risk decisions local while escalating high-risk or large-context reasoning to edge/cloud resources.

\begin{figure*}[t]
\centering
\includegraphics[width=0.88\textwidth]{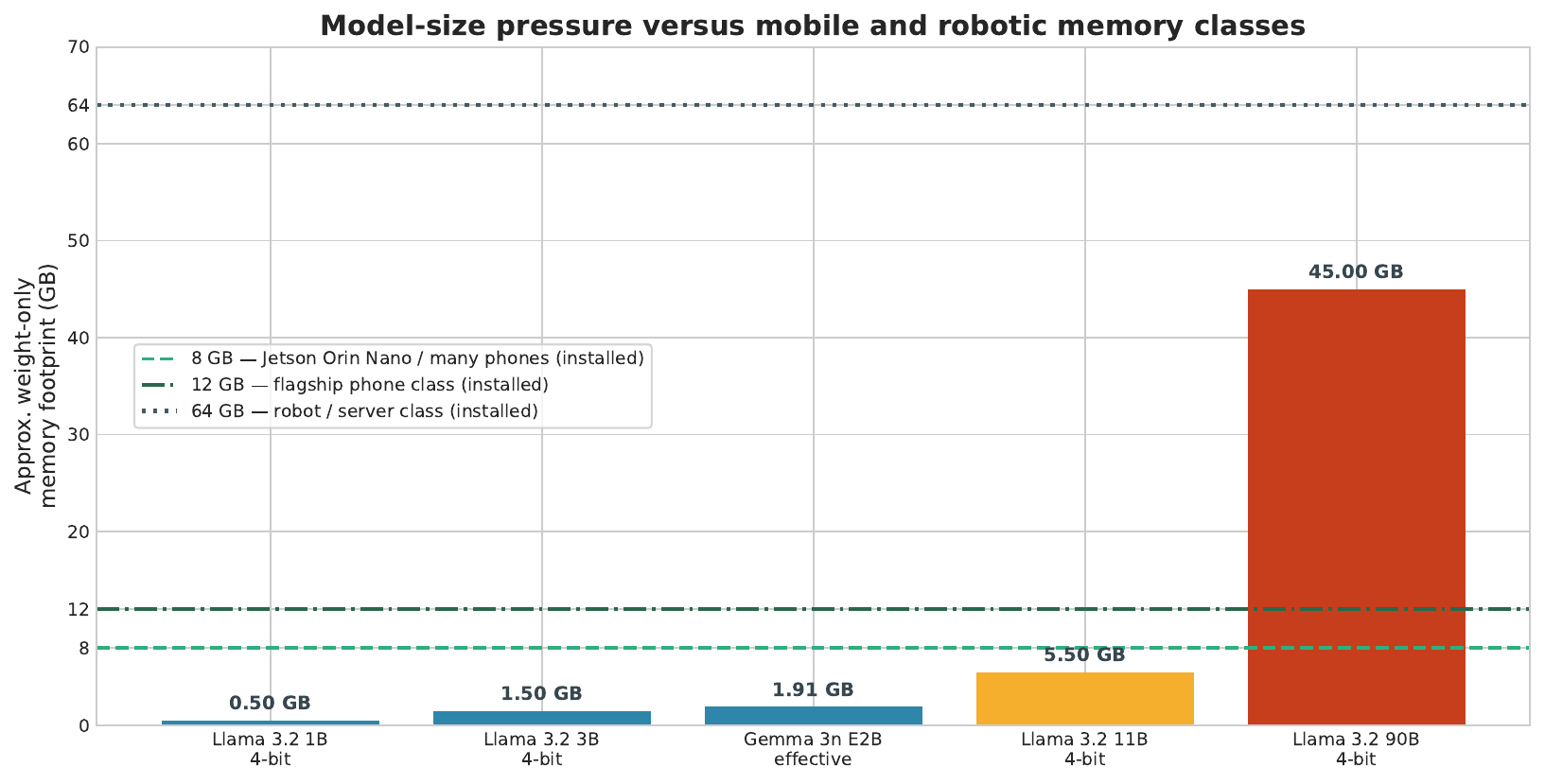}
\caption{Approximate model-size pressure relative to mobile and robotic memory classes. Bars show weight-only memory estimates derived from parameter count and quantization assumptions, whereas horizontal lines show representative 8~GB, 12~GB, and 64~GB device-memory classes. The final deployed footprint is larger because it includes OS memory, foreground apps, sensor pipelines, KV cache, activations, model runtimes, and thermal throttling; therefore the figure supports conservative placement of 1B–3B class models on phones/robot edge boxes, compact sensing models on watches/glasses, and larger VLMs on robot/server-class resources.}
\label{fig:model_memory_fit}
\end{figure*}

Figure~\ref{fig:model_memory_fit} makes this deployment boundary explicit. A 1B model has an approximate 4-bit weight-only footprint of 0.5~GB and a 3B model about 1.5~GB before runtime overhead, making them plausible on smartphones and robot edge boxes. Gemma~3n E2B's effective 1.91B memory load falls in a similar mobile-class range \cite{google2025gemma3n}. In contrast, Llama~3.2 11B vision has an approximate 4-bit weight-only footprint of 5.5~GB, which is more realistic on robot/edge platforms than on heavily loaded phones, and the 90B vision model is roughly 45~GB even before activations, KV cache, operating-system memory, and concurrent applications \cite{meta2024llama32,meta2024llama32card}. This supports a staged architecture: mobile and wearable devices sense, summarize, and trigger; robot or edge-class devices maintain richer local world models; and cloud resources handle very large VLM/LLM reasoning when privacy, latency, and connectivity permit.

\subsection{Robotics and Physical Actuation}

Another enabler is the progress of robotics and actuation technologies. While many proactive systems provide information or recommendations, emerging systems may also physically act in the environment. Robots, smart home devices, assistive systems, and autonomous vehicles can directly affect the physical world. The industrial base for this transition is already large: the International Federation of Robotics reports that 542,000 industrial robots were installed in 2024, annual installations exceeded 500,000 units for the fourth consecutive year, and the operational stock reached 4.664 million robots worldwide \cite{ifr2025worldrobotics}. This does not mean that all robots are proactive systems, but it shows that physical actuation is becoming a sufficiently widespread computational substrate for proactive services.

This expands proactive computing from information assistance to physical assistance. For example, a robot may help an older adult before a fall risk increases, a smart home may adjust lighting and temperature before discomfort occurs, or an autonomous vehicle may intervene before a dangerous situation develops.

Physical actuation increases the value of proactive computing, but it also increases risk. Incorrect physical actions can have safety consequences. Thus, proactive systems involving actuation require stronger reliability, transparency, user control, and safety guarantees.

\begin{table*}[t]
\centering
\caption{Technology landscape for proactive computing enablers. The table lists the main technologies discussed in this section from broad classes to concrete subtechnologies, together with current industrial and research status. Section~\ref{sec:technical_challenges} should cover the corresponding technical risks: semantic grounding, personalization, prediction-to-action decision making, uncertainty, latency, energy, privacy, and evaluation.}
\label{tab:enabler_landscape}
\footnotesize
\begin{tabularx}{\textwidth}{p{0.13\textwidth} p{0.17\textwidth} X X X}
\toprule
\textbf{Enabler} & \textbf{Subtechnologies} & \textbf{Industrial status} & \textbf{Research status and numbers} & \textbf{Why it matters for proactivity} \\
\midrule
Personal and pervasive sensing & Smartphones, smartwatches, smart rings, earbuds, mobile cameras, inertial sensors, location, physiological sensors, neural time-series & Wearables are a mass-market substrate, with 611.5 million shipments in 2025 and a 2030 projection of 689.7 million units \cite{idc2026wearables}. & Wearable and neural sensing research has shifted toward multimodal and high-resolution representation learning, targeting robust generalization across diverse users \cite{zhang2022har,willett2021handwriting}. & Provides continuous behavioral, physiological, and cognitive signals for detecting availability, stress, routine, and intent. \\
\addlinespace
Ambient and IoT sensing & Smart-home sensors, cameras, microphones, environmental sensors, Wi-Fi/Bluetooth/cellular IoT, building and city infrastructure & Connected IoT devices are estimated to grow from 18.5 billion in 2024 to 21.1 billion in 2025 and 39 billion in 2030; Wi-Fi, Bluetooth, and cellular represent 32\%, 24\%, and 22\% of 2025 IoT connections \cite{iotanalytics2025connections}. & Research focuses on context-aware computing, sensor fusion, privacy-aware sensing, and robust inference from heterogeneous streams \cite{dey2001understanding,zhang2022har}. & Extends proactivity from personal devices to rooms, buildings, mobility systems, and public environments. \\
\addlinespace
AI-based context understanding & HAR, representation learning, multimodal learning, self-supervised learning, foundation models, uncertainty estimation & Foundation-model capabilities are being integrated into consumer assistants, operating systems, robotics stacks, and edge/cloud AI services, but safety and grounding are still deployment constraints. & GPT-3 scaled few-shot language modeling to 175 billion parameters, CLIP trained on 400 million image–text pairs and evaluated zero-shot transfer on over 30 datasets, and foundation-model surveys identify broad opportunities and risks \cite{brown2020language,radford2021learning,bommasani2021foundation}. & Converts raw signals into semantic state, predicts future needs, and supports decision policies over possible actions. \\
\addlinespace
Distributed and edge infrastructure & On-device inference, split computing, cloudlets, edge servers, 5G/6G, Open RAN, federated learning & Commercial AI services increasingly combine on-device inference with nearby edge or cloud resources; this is necessary because mobile and wearable devices remain battery-, memory-, and thermally constrained. & Edge computing targets low-latency operation on the order of a few tens of milliseconds, and Neurosurgeon reported a 3.1\(\times\) average latency improvement through DNN partitioning \cite{satyanarayanan2017edge,kang2017}. Federated and secure aggregation methods provide decentralized personalization primitives \cite{mcmahan2017communication,bonawitz2017practical}. & Determines whether proactive inference can be timely, private, personalized, and energy efficient. \\
\addlinespace
Device capability and model fit & Smartphone NPUs, wearable neural engines, smart-glasses SoCs, robot edge boxes, quantized LLMs, compact VLM/encoder models, model routing & Flagship phones provide mobile-class memory and NPUs (e.g., 12~GB RAM on Galaxy S25 Ultra and a 16-core Neural Engine on iPhone 16 Pro), watches/glasses expose sensors with tighter batteries, and Jetson Orin Nano Super provides 8~GB memory and 67~INT8 TOPS for robot/edge systems \cite{samsung2025galaxys25,apple2024iphone16pro,apple2024watchseries10,meta2023rayban,nvidia2024orinnanosuper}. & Llama~3.2 1B/3B and Gemma~3n show that pruning, distillation, quantization, PLE caching, MatFormer activation, and conditional parameter loading can move 1B--4B class models toward mobile deployment; larger 11B/90B vision models remain robot/server-class in practice \cite{meta2024llama32,meta2024llama32card,google2025gemma3n}. & Determines which proactive functions can be local, which need phone/edge assistance, and which require cloud-scale reasoning. \\
\addlinespace
Robotics and physical actuation & Service robots, industrial robots, assistive robots, smart-home actuators, autonomous vehicles, drones & The International Federation of Robotics reports 542,000 industrial robot installations in 2024 and 4.664 million operational industrial robots worldwide \cite{ifr2025worldrobotics}. & Robotics research increasingly combines perception, planning, reinforcement learning, embodied AI, and language-conditioned control, but safety validation remains more demanding than for recommendation-only systems \cite{shah22lmnav,yokoyama2024vlfm}. & Turns proactivity from notification into physical intervention, increasing both usefulness and safety risk. \\
\bottomrule
\end{tabularx}
\end{table*}

\section{Design Space of Proactive Computing}
\label{sec:design_space}

\subsection{Sensing Modality and Granularity}

The first design dimension is sensing. 
Proactive systems differ in the types of signals they observe and the granularity at which they collect them. 
Some systems rely on a single modality, such as location or activity, while others integrate multimodal data from wearables, smartphones, environmental sensors, cameras, and infrastructure.

Granularity is also important. 
Continuous high-resolution sensing can improve context understanding, but it increases energy consumption and privacy risks. 
In contrast, sparse or event-driven sensing can reduce overhead, but may miss important contextual changes. 
This trade-off becomes more important as wearables move from episodic interaction toward always-available physiological and behavioral sensing at hundreds of millions of deployed units per year \cite{idc2026wearables}. 
Therefore, proactive systems must balance sensing fidelity against resource cost and privacy exposure.

\subsection{Context and Intent Understanding}

The second design dimension is the level of understanding. 
Simple systems may classify low-level context, such as location or activity. 
More advanced systems may infer high-level states such as stress, attention, availability, intent, preference, or risk.

For proactive computing, high-level understanding is essential because actions are rarely determined by raw context alone. 
The same activity may imply different needs depending on the user, time, environment, and history. 
Thus, proactive systems require semantic and personalized interpretation of context.

\subsection{Prediction Horizon}

The third dimension is prediction horizon. 
Some proactive systems operate over short horizons, such as predicting immediate interruption risk or next activity. 
Others operate over longer horizons, such as predicting health deterioration, long-term behavior patterns, or future mobility needs.

Short-horizon prediction often supports real-time intervention, while long-horizon prediction supports planning and prevention. 
The appropriate horizon depends on the application domain and the type of action being taken.

\subsection{Personalization Level}

The fourth dimension is personalization. 
Proactive computing is inherently personal because the usefulness and acceptability of actions depend on individual preferences, routines, and values. 
A generic proactive action may be useful for one user but annoying or inappropriate for another.

Personalization can occur at multiple levels:
\begin{itemize}
    \item user-specific sensing calibration;
    \item personalized activity or context models;
    \item preference learning for recommendations;
    \item adaptive intervention timing;
    \item user-specific autonomy settings.
\end{itemize}

However, personalization requires data. 
This creates tension between personalization quality and privacy preservation. 
Local learning, federated learning, and privacy-preserving adaptation are therefore important directions. Federated learning demonstrated that models can be trained across mobile devices while keeping raw training data decentralized \cite{mcmahan2017communication}; differential privacy provides a formal framework for limiting disclosure from statistical computation \cite{dwork2006calibrating}; and secure aggregation enables a server to compute aggregate model updates without inspecting individual device updates \cite{bonawitz2017practical}.

\subsection{Autonomy Level}

The fifth dimension is autonomy. 
Not all proactive systems should act with the same degree of independence. 
A system may simply notify the user, recommend an action, ask for confirmation, automatically execute a digital action, or physically intervene in the environment.

We can organize autonomy into the following levels:
\begin{itemize}
    \item \textbf{Information}: the system provides relevant information.
    \item \textbf{Recommendation}: the system suggests an action.
    \item \textbf{Confirmation-based automation}: the system asks for approval before acting.
    \item \textbf{Automatic digital action}: the system executes software-level actions.
    \item \textbf{Physical actuation}: the system acts in the physical environment.
\end{itemize}

Higher autonomy may increase convenience, but it also increases risk and requires stronger trust, accountability, and user control.

\subsection{Deployment Architecture}

The sixth dimension is deployment. 
Proactive systems may be deployed on-device, at the edge, in the cloud, or across hybrid architectures. 
On-device deployment provides low latency and stronger privacy but is constrained by resources. 
Cloud deployment provides high computational capacity but increases latency and privacy concerns. 
Edge deployment can provide a middle ground, especially for latency-sensitive and resource-intensive services. For DNN inference, collaborative device-edge execution can materially change performance: Neurosurgeon reports a 3.1\(\times\) average latency improvement by partitioning computation across mobile and cloud resources \cite{kang2017}.

\begin{figure*}[t]
\centering
\includegraphics[width=\textwidth]{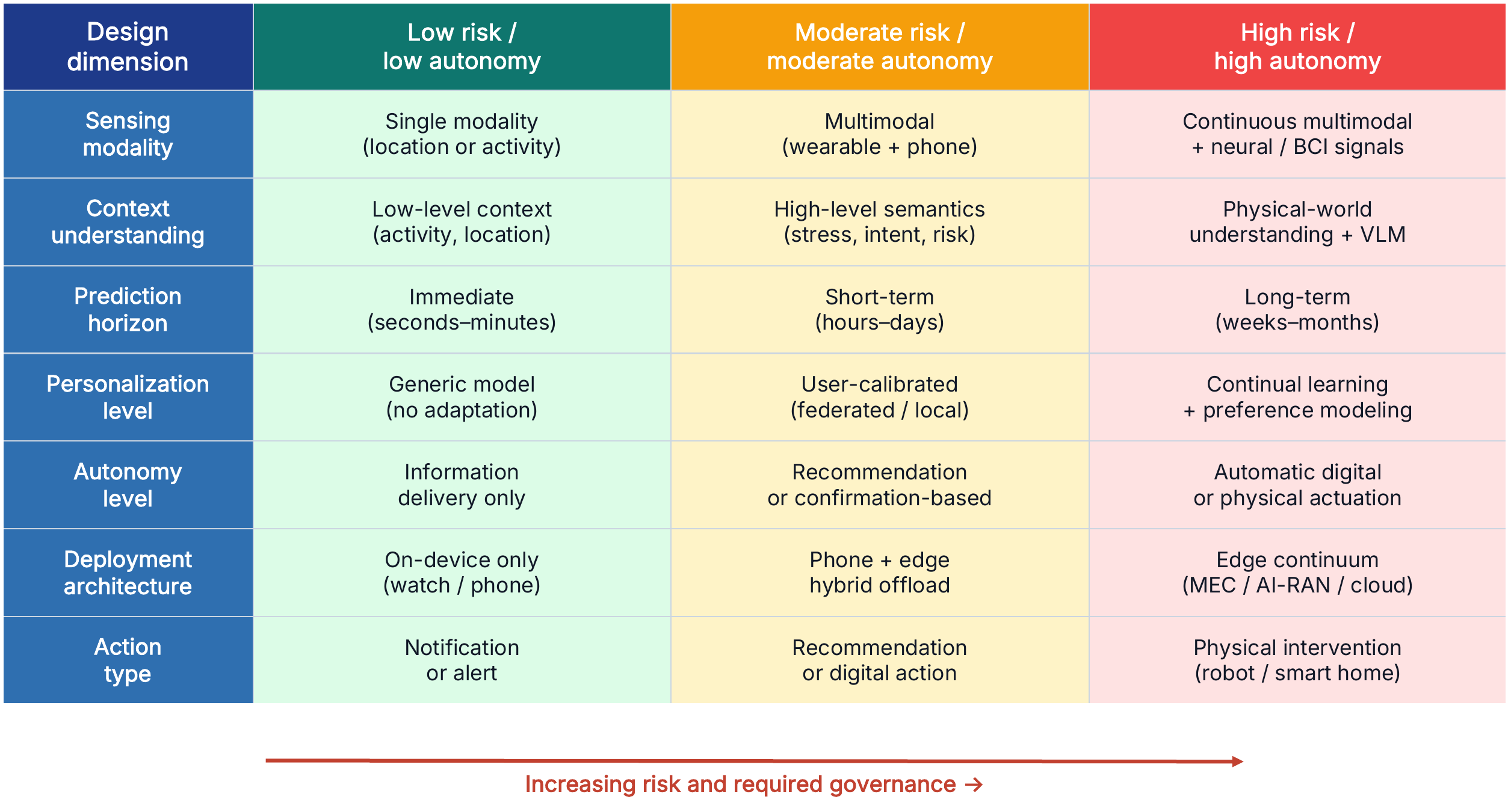}
\caption{Design-space matrix for proactive computing. The figure maps sensing modality, context understanding, prediction horizon, personalization, autonomy level, deployment architecture, and action type against increasing risk and required governance.}
\label{fig:design_space_matrix}
\end{figure*}

The deployment architecture should be chosen based on application requirements, including latency, energy, privacy, model complexity, and communication availability. Figure~\ref{fig:design_space_matrix} is a suggested design-space figure to make these trade-offs visually explicit.

\subsection{Action Type}

The final dimension is action. 
Actions can range from low-risk information delivery to high-risk physical intervention. 
This dimension is central to proactive computing because proactivity is defined not only by prediction but also by action.

Typical action types include:
\begin{itemize}
    \item proactive notification;
    \item recommendation;
    \item scheduling or preparation;
    \item environmental control;
    \item health or safety intervention;
    \item robotic assistance;
    \item autonomous control.
\end{itemize}

Each action type requires different levels of reliability, explanation, user consent, and evaluation.

\section{Technical Challenges and Research Trends}
\label{sec:technical_challenges}

Section~\ref{sec:enablers} introduced a set of concrete technology families: personal and wearable sensing, ambient IoT sensing, BCI and neurotechnology, AI-based context understanding, distributed edge infrastructure, HAR/time-series and physical context understanding, device capability and on-device model fit, and robotics or physical actuation. The original challenge structure of this section already covered several cross-cutting issues, especially semantic inference, personalization, latency, energy, uncertainty, and evaluation. However, BCI-specific neural privacy, ambient-scale deployment heterogeneity, local model memory pressure, and safety validation for physical actuation require more explicit treatment. Table~\ref{tab:enabler_challenge_mapping} therefore maps each enabler to the technical challenges that must be addressed before it can support reliable proactive computing.

\begin{table*}[t]
\centering
\caption{Coverage of Section~\ref{sec:enablers} enablers in Section~\ref{sec:technical_challenges}. The table shows that Section~\ref{sec:technical_challenges} covers the major challenge classes, but also makes explicit where additional BCI-, IoT-, and actuation-specific validation is needed.}
\label{tab:enabler_challenge_mapping}
\footnotesize
\begin{tabularx}{\textwidth}{p{0.17\textwidth} X X X}
\toprule
\textbf{Enabler from Section~\ref{sec:enablers}} & \textbf{Main technical issue} & \textbf{Evidence and quantitative anchor} & \textbf{Coverage in this section} \\
\midrule
Personal and wearable sensing & Robust inference across users, devices, and activities & Wearable shipments reached 611.5 million units in 2025, but HAR surveys identify cross-user and cross-device robustness as a persistent research problem \cite{idc2026wearables,zhang2022har}. & Covered by semantic inference, personalization, and evaluation; additional benchmark standardization remains needed. \\
\addlinespace
Ambient and IoT sensing & Heterogeneous sensing, missing data, network dependency, and privacy in shared spaces & Connected IoT devices are projected to rise from 21.1 billion in 2025 to 39 billion in 2030, creating a large but fragmented sensing substrate \cite{iotanalytics2025connections}. & Covered by semantic inference and resource constraints; privacy and shared-space consent connect to Section~\ref{sec:socio_technical}. \\
\addlinespace
BCI and neurotechnology & Neural-signal robustness, calibration, safety, and neuroprivacy & Research BCIs report 90 characters/min for handwriting and 62 words/min for attempted speech, while the SWITCH trial reported 12-month safety follow-up in four participants \cite{willett2021handwriting,willett2023speech,mitchell2023stentrode}. & Partially covered by uncertainty and evaluation; now explicitly linked to neural privacy, calibration, and high-stakes failure handling. \\
\addlinespace
AI-based context understanding & Grounding, hallucination, time-series transfer, physical reasoning, uncertainty estimation, and prediction-to-action decision making & GPT-3 used 175 billion parameters and CLIP used 400 million image--text pairs, but HAR/time-series surveys, PhysBench, and hallucination studies show that scale does not eliminate grounding risk or physical-context gaps \cite{brown2020language,radford2021learning,bommasani2021foundation,bai2024hallucination,yuan2024selfsupervisedhar,zhang2024sslts,chow2025physbench}. & Covered by semantic inference, uncertainty-aware proactivity, and prediction-to-action gap; now explicitly linked to HAR, time-series, adaptive sensing, and physical-world understanding. \\
\addlinespace
Distributed and edge infrastructure & Latency, partitioning, energy, privacy-preserving personalization, on-premise governance, AI-RAN placement, and network failure & Edge systems target latencies of a few tens of milliseconds, Neurosurgeon reported a 3.1\(\times\) average latency improvement from DNN partitioning, ETSI MEC standardizes network-edge context exposure, and vendor-reported AI-RAN trials show base-station compute co-location \cite{satyanarayanan2017edge,kang2017,etsi2018mec5g,nvidia2024airan}. & Covered by latency, energy, and personalization; now framed as an edge-continuum placement and orchestration problem requiring end-to-end testing under mobility, congestion, and administrative boundaries. \\
\addlinespace
Device capability and on-device model fit & Free RAM, thermal throttling, NPU/GPU scheduling, model quantization, and staged offloading & Flagship phones can offer 12~GB RAM or dedicated neural engines, but watches/glasses are tighter and larger VLMs such as Llama~3.2 11B/90B exceed conservative local mobile envelopes without offload \cite{samsung2025galaxys25,apple2024iphone16pro,meta2024llama32,meta2024llama32card}. & Covered by latency, energy, and resource constraints; now explicitly linked to installed-versus-free RAM, KV cache, foreground apps, and split execution. \\
\addlinespace
Robotics and physical actuation & Safety validation, reversibility, accountability, and human override & Industrial robot installations reached 542,000 units in 2024, with 4.664 million robots in operation worldwide \cite{ifr2025worldrobotics}. & Partially covered by prediction-to-action and evaluation; safety cases and physical-world failure recovery need stronger domain-specific evaluation. \\
\bottomrule
\end{tabularx}
\end{table*}

\subsection{From Low-Level Sensing to High-Level Semantics}

A fundamental challenge is converting low-level sensor signals into high-level semantic understanding. 
Many sensing systems can classify activities or detect events, but proactive computing requires deeper interpretation. 
For example, knowing that a user is walking does not directly indicate whether the user needs navigation assistance, health monitoring, interruption avoidance, or no intervention at all.

Current research addresses this challenge through multimodal learning, self-supervised representation learning, sensor fusion, and foundation-model-based reasoning \cite{bommasani2021foundation}. 
The goal is to build representations that are robust across users, devices, environments, and tasks. For BCI-enabled proactivity, the semantic gap is even narrower and more safety-critical: neural signals may support high-bandwidth communication, such as 90 characters/min handwriting decoding or 62 words/min attempted-speech decoding, but proactive systems must distinguish intentional control, background neural variation, fatigue, and calibration drift before initiating an action \cite{willett2021handwriting,willett2023speech}. 
Because multimodal large language models may generate outputs that are not grounded in the observed evidence, proactive systems also require verification and confidence checks before model outputs are translated into actions \cite{bai2024hallucination}.

\subsection{Personalization and Cold Start}

Proactive services must adapt to individual users. 
However, personalization is difficult when user data are limited, noisy, or sensitive. 
Cold-start users may not provide enough data for reliable adaptation, while long-term users may change behavior over time.

Research trends include federated learning, continual learning, meta-learning, and preference learning. 
These methods aim to personalize models without requiring centralized collection of sensitive data. 
Federated averaging was introduced precisely for communication-efficient learning from decentralized mobile data, while secure aggregation and differential privacy provide complementary protections for individual updates and statistical outputs \cite{mcmahan2017communication,bonawitz2017practical,dwork2006calibrating}. 
However, balancing personalization, privacy, and robustness remains an open challenge.

\subsection{The Prediction-to-Action Gap}

Perhaps the most important technical gap is the prediction-to-action gap. 
Many systems can predict future states, but deciding whether and how to act is more difficult. 
An accurate prediction does not automatically imply that an intervention is needed.

For example, a system may predict that a user is likely to be stressed, but it must still decide whether to notify the user, recommend rest, adjust the environment, delay intervention, or do nothing. 
The correct action depends on user preference, uncertainty, context, and potential harm.

Emerging directions include reinforcement learning, planning-based decision making, risk-sensitive policies, and human-in-the-loop interaction. 
These approaches shift the focus from prediction accuracy to action quality. The shift is especially important when proactivity has physical consequences. In robotics and smart environments, a wrong intervention can change the state of the physical world; therefore, action policies should encode reversibility, safe fallback states, explicit user override, and conservative thresholds before actuation. The industrial scale of robotics---542,000 new installations in 2024 and 4.664 million operational industrial robots worldwide---makes this safety issue practically important rather than purely speculative \cite{ifr2025worldrobotics}.

\subsection{Uncertainty-Aware Proactivity}

Proactive systems must operate under uncertainty. 
Sensor measurements may be noisy, models may be wrong, and user preferences may be ambiguous. 
False interventions can reduce user trust and cause annoyance.

Therefore, proactive systems require uncertainty-aware mechanisms. 
These include confidence estimation, threshold-based triggering, Bayesian modeling, selective prediction, fallback strategies, and conservative action policies. 
Bayesian approximation methods such as Monte Carlo dropout and ensemble-based uncertainty estimation provide practical examples of how predictive uncertainty can be exposed to downstream decision policies \cite{gal2016dropout,lakshminarayanan2017simple}. 
In many cases, the most important capability is not acting, but knowing when not to act.

\subsection{Latency, Energy, and Resource Constraints}

Proactive computing often requires continuous sensing and timely inference. 
This creates significant resource challenges, especially on mobile and wearable devices. 
Always-on sensing and AI inference can drain battery, increase thermal load, and reduce device usability.

Current research explores adaptive sampling, event-driven sensing, model compression, pruning, quantization, early-exit models, tinyML, and edge offloading. 
These techniques aim to make proactive computing efficient enough for long-term deployment. For edge-continuum deployments, the open technical problem is no longer a one-time choice between cloud and device. The system must dynamically decide whether to execute a step on a wearable, phone, on-premise server, MEC/base-station AI node, or regional cloud while respecting latency, privacy, cost, energy, radio condition, and model uncertainty. On-premise edge can improve privacy and user experience when sensitive data should remain inside a building or organization; MEC and AI-RAN can improve responsiveness when compute is placed near the radio path; and cloud resources remain useful for large-context reasoning. However, these benefits depend on tail latency, mobility, congestion, service handoff, failure recovery, and verifiable data-governance boundaries rather than on average latency alone.
Model compression is particularly relevant: deep compression showed that neural networks can be reduced by 35\(\times\) to 49\(\times\) without accuracy loss on representative vision models, making on-device inference more practical \cite{han2015deep}. More recent mobile foundation-model deployments extend the same principle through quantization, pruning, distillation, caching, early exiting, and conditional parameter loading. For example, Llama~3.2 1B and 3B are positioned for selected edge and mobile devices, while Gemma~3n reduces E2B's effective memory load to 1.91B parameters through PLE caching and selective parameter activation \cite{meta2024llama32,google2025gemma3n}. Nevertheless, installed RAM is not the same as usable model memory: the operating system, foreground apps, sensor pipelines, KV cache, activations, and thermal governors reduce the practical envelope. This is why watches and smart glasses should primarily run compact sensing and interaction models, phones should run small quantized LLMs or local routing policies, robot edge boxes can host richer perception/planning loops, and large VLM/LLM reasoning should remain edge/cloud-assisted when privacy and latency constraints allow. 
However, efficiency must also be evaluated environmentally; Green AI argues that model quality should be assessed together with computational cost, and prior work has shown that large neural model training can impose substantial energy and carbon costs \cite{schwartz2020green,strubell2019energy}.

\subsection{Evaluation Challenges}

Evaluating proactive computing is difficult because conventional metrics such as accuracy or F1 score are insufficient. 
A proactive action must be timely, useful, acceptable, and non-intrusive. 
Moreover, the value of an action often depends on long-term user experience rather than immediate prediction correctness.

Potential evaluation metrics include:
\begin{itemize}
    \item action precision and false intervention rate;
    \item user acceptance rate;
    \item perceived usefulness;
    \item annoyance or interruption cost;
    \item trust and transparency;
    \item long-term retention;
    \item energy and sustainability cost.
\end{itemize}

\begin{figure}[t]
\centering
\includegraphics[width=0.9\columnwidth]{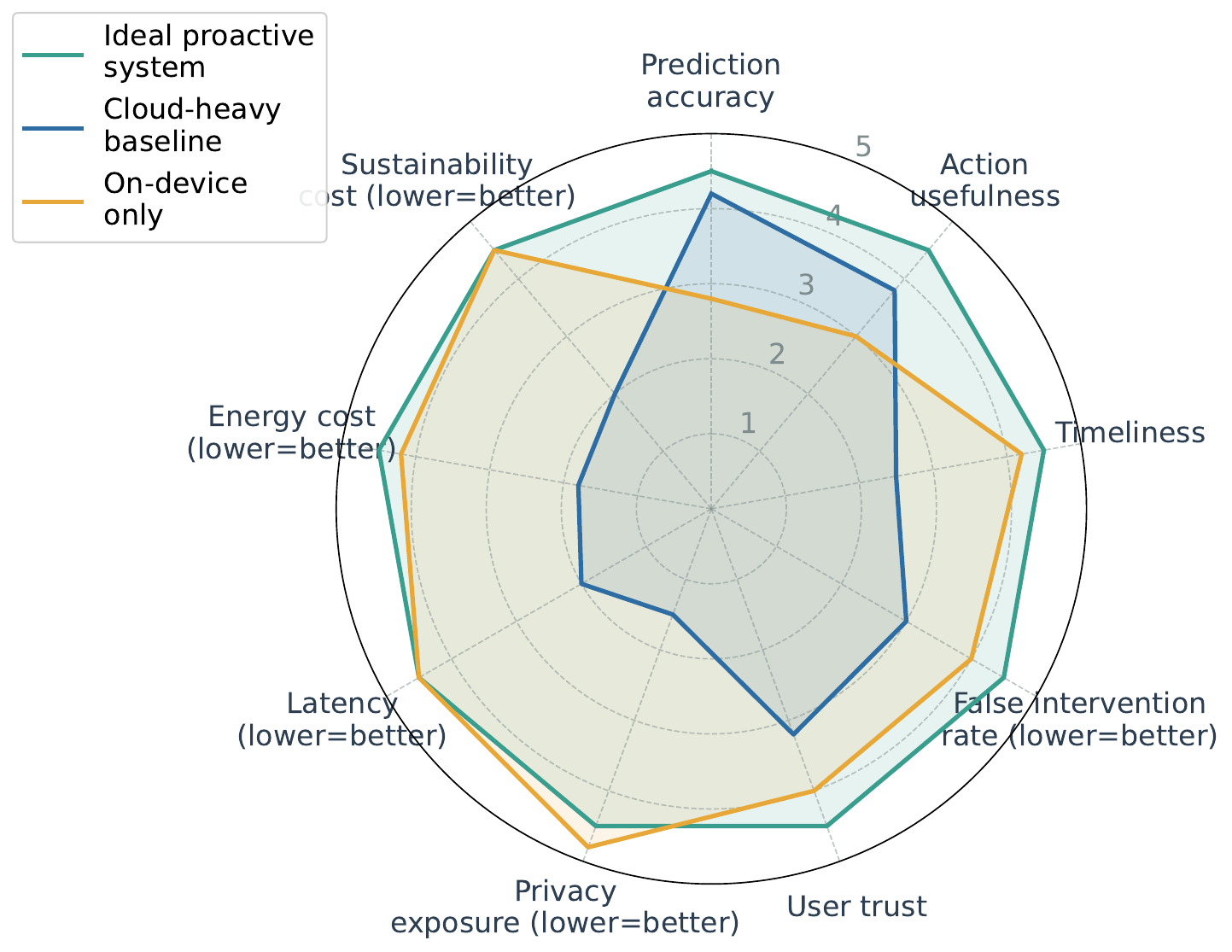}
\caption{Multi-objective evaluation view for proactive computing. The figure emphasizes that proactive systems cannot be evaluated only by prediction accuracy; they also require action-level, human-centered, privacy, latency, energy, and sustainability metrics.}
\label{fig:evaluation_metrics_radar}
\end{figure}

Developing standardized benchmarks and evaluation protocols remains an important open problem. For HAR and time-series proactive systems, evaluation should report cross-user transfer, cross-device transfer, temporal calibration, missing-sensor robustness, long-horizon forecasting quality, and false-intervention cost rather than only short-window F1. For physical-world understanding, PhysBench shows that VLMs still struggle across object properties, object relationships, scene understanding, and physics-based dynamics, so evaluation must include physically grounded failure modes before proactive systems are allowed to actuate \cite{chow2025physbench}. For BCI-enabled proactive systems, evaluation should include neural-signal calibration time, long-term signal stability, false activation rate, user-perceived agency, and neuroprivacy exposure in addition to task performance; for example, the SWITCH study reported stability over 12 months in four participants, whereas high-speed decoding studies report communication rates but still require longitudinal validation before everyday deployment \cite{mitchell2023stentrode,willett2021handwriting,willett2023speech}. Figure~\ref{fig:evaluation_metrics_radar} suggests a figure that can summarize this broader evaluation space.

\section{Socio-Technical Challenges}
\label{sec:socio_technical}

\subsection{User Acceptance}

Unlike conventional systems, proactive systems act before explicit user requests. 
This can improve convenience, but it can also create surprise, discomfort, or resistance. 
A technically correct action may still be rejected if the user perceives it as unnecessary, intrusive, or poorly timed.

User acceptance depends on several factors, including perceived usefulness, timing, explanation, controllability, and cultural norms. Technology acceptance research has long shown that perceived usefulness and perceived ease of use shape adoption intentions \cite{davis1989perceived}, and UTAUT further identifies performance expectancy, effort expectancy, social influence, and facilitating conditions as determinants of acceptance \cite{venkatesh2003user}. 
Therefore, proactive systems should provide mechanisms for user feedback, adjustable autonomy, and personalization. 
The design goal is not simply to maximize action frequency, but to maximize appropriate and accepted interventions.

Recent survey evidence provides a concrete rationale for this design direction. In Microsoft and LinkedIn's 2024 Work Trend Index survey of 31,000 people across 31 countries, 75\% of knowledge workers reported using AI at work and 78\% of AI users reported bringing their own AI tools, suggesting that users adopt AI when it reduces friction faster than formal organizational workflows can adapt \cite{microsoft2024worktrend}. Adecco's 2024 Global Workforce of the Future survey of 35,000 workers across 27 economies reports that AI users save an average of one hour per working day, with 20\% saving up to two hours and 5\% saving three to four hours; the same report states that users reinvest saved time into creative work (28\%), strategic thinking (26\%), and work--life balance (27\%) \cite{adecco2024globalworkforce}. McKinsey's 2025 workplace AI survey similarly reports that 92\% of companies plan to increase AI investment while only 1\% describe their deployment as mature, indicating a large gap between perceived value and operational integration \cite{mckinsey2025superagency}. These findings support the core premise of proactive computing: users do not merely want faster interfaces; they want computational systems that reduce avoidable labor, anticipate routine needs, and free attention for higher-value human goals.

At the same time, survey evidence also warns against an unqualified automation narrative. Pew Research Center reports that U.S. workers are more worried than hopeful about future workplace AI use, and Stanford HAI's discussion of worker preferences emphasizes that people often prefer AI support for repetitive or low-value tasks while resisting automation that removes agency or affects identity-laden work \cite{pew2025aiworkplace,stanfordhai2024workerwants}. Thus, the correct implication is not full autonomy, but human-centered proactivity: systems should infer wants and handle low-risk, high-friction work in advance, while preserving confirmation, explanation, and override for consequential decisions.

\begin{table*}[t]
\centering
\caption{Survey evidence connecting AI use to time savings, perceived usefulness, and user demand for agency-preserving automation. The table summarizes large-scale public reports from Adecco, Microsoft/LinkedIn, McKinsey, and Pew. It supports the paper's normative claim that proactive computing is valuable when it reduces avoidable labor and anticipates routine needs, but must remain controllable because users are not uniformly comfortable with unrestricted automation.}
\label{tab:ai_user_value_survey}
\footnotesize
\begin{tabularx}{\textwidth}{>{\columncolor{gray!20}\bfseries}p{0.18\textwidth} p{0.18\textwidth} X X}
\toprule
\rowcolor{gray!50}
Survey \& Scope & Primary finding & Key metrics & Implication for proactive AI \\
\midrule
Microsoft \& LinkedIn 2024 Work Trend Index\newline (31,000 workers, 31 countries) & Users adopt AI rapidly to bypass organizational friction & 75\% of knowledge workers use AI at work; 78\% of AI users bring their own AI tools & Proactive systems will be adopted if they genuinely reduce friction, even outside formal IT channels \\
\addlinespace
Adecco 2024 Global Workforce of the Future\newline (35,000 workers, 27 economies) & AI generates measurable daily time savings & Average 1 hour saved per day; 20\% save up to 2 hours; 5\% save 3--4 hours & The value of proactivity is reclaiming time for creative work, strategy, and work-life balance \\
\addlinespace
McKinsey 2025 Workplace AI Survey & High expected value but low operational maturity & 92\% of companies plan to increase AI investment; only 1\% report mature deployment & The gap between value and deployment maturity highlights the difficulty of system integration \\
\addlinespace
Pew Research Center \& Stanford HAI (U.S. workers) & Workers want support for routine tasks, not full replacement & Workers are more worried than hopeful; prefer automation of repetitive, low-value tasks & Proactive systems must preserve human agency and avoid automating identity-laden or high-stakes decisions \\
\bottomrule
\end{tabularx}
\end{table*}

\subsection{Trust and Reliability}

Trust is central to proactive computing. In automation research, trust is commonly treated as an attitude that an automated agent will help achieve an individual's goals under uncertainty and vulnerability \cite{lee2004trust}. 
Because proactive systems initiate actions without explicit requests, users must believe that the system understands the situation and acts in their interest. 
Incorrect or poorly timed interventions can rapidly erode trust.

Reliability must therefore be treated as a first-class design constraint. 
Systems should estimate uncertainty, avoid high-risk actions under low confidence, and provide explanations for their behavior. Trust should also be measured empirically: Kohn et al. review 116 trust-in-automation measurement approaches, showing that trust is not a single scalar but a construct that can be assessed through questionnaires, behavioral measures, and physiological indicators \cite{kohn2021measurement}. 
In many cases, conservative behavior is preferable to excessive automation.

A key principle is that proactive systems must learn when not to act. 
Avoiding unnecessary interventions is as important as providing useful ones.

\subsection{Privacy and Data Governance}

Proactive computing relies on highly personal data, including location, activity, physiology, routines, preferences, and social context. 
Continuous sensing can create a perception of surveillance, especially when data are transmitted to external servers.

Privacy-preserving architectures are therefore essential. 
Potential approaches include on-device processing, federated learning, differential privacy, secure aggregation, and data minimization \cite{mcmahan2017communication,dwork2006calibrating,bonawitz2017practical}. 
Rather than collecting all available data, proactive systems should collect and process only the data necessary for a given service.

Data governance is also important. 
Users should understand what data are collected, where they are processed, how long they are stored, and how they influence system actions.

\subsection{Policy and Accountability}

Proactive computing raises policy and accountability questions. 
If a system acts before user input and causes harm, it may be unclear who is responsible. 
Responsibility may involve device manufacturers, model developers, service providers, infrastructure operators, or users.

These questions become more complex in distributed infrastructures such as edge computing, 6G, and Open RAN. 
Data and computation may move across devices, networks, regions, and organizations. 
Thus, proactive computing systems require governance mechanisms that clarify responsibility, data ownership, auditability, and regulatory compliance. The NIST AI Risk Management Framework identifies trustworthy AI characteristics including valid and reliable, safe, secure and resilient, accountable and transparent, explainable and interpretable, privacy-enhanced, and fair with harmful bias managed, providing a useful governance checklist for proactive systems \cite{nist2023airmf}.

\subsection{Socio-Cultural and Ethical Issues}

The acceptability of proactive behavior is culturally dependent. 
Some users or societies may value convenience and automation, while others may prioritize autonomy and privacy. 
A proactive action that is perceived as helpful in one context may be perceived as intrusive in another.

Ethical issues also arise when proactive systems influence user behavior. 
Systems may nudge users toward healthier, safer, or more efficient actions, but such nudging can become manipulative if users lack awareness or control. 
Therefore, proactive systems should preserve user agency through transparency, consent, and override mechanisms.

Fairness is another concern. 
Models trained on limited populations may perform poorly for users with different bodies, behaviors, cultures, environments, or device usage patterns. 
Proactive systems must therefore consider fairness and inclusiveness during design and evaluation.

\subsection{Sustainability}

Sustainability is an important but often underexplored issue in proactive computing. 
Continuous sensing, always-on inference, edge communication, and large AI models can increase energy consumption and carbon cost. Prior work on Green AI argues that accuracy should be reported alongside computational cost, while energy analysis of neural NLP models shows that model development and training choices can have substantial environmental impacts \cite{schwartz2020green,strubell2019energy}. 
This is particularly problematic for wearable and IoT devices, where battery capacity is limited.

Sustainable proactive computing should avoid unnecessary sensing and computation. 
Promising directions include event-driven sensing, adaptive sampling, duty cycling, selective computation, model compression, and carbon-aware scheduling. 
A proactive system should not simply be always on; it should wake up and compute only when doing so is likely to provide value.

\section{Future Research Directions}
\label{sec:future}

\subsection{Local-First Proactive Computing}

Future proactive systems should prioritize local processing whenever possible. 
Local-first architectures can reduce latency, preserve privacy, and improve robustness under network variability. 
However, local devices have limited resources, so efficient models and adaptive computation mechanisms are required.

A promising direction is hybrid local-edge intelligence, where lightweight models perform continuous local monitoring and selectively invoke on-premise, MEC/base-station AI, or cloud resources for complex reasoning. This direction is supported by edge-computing work that targets low-latency offload services, by empirical split-inference results showing 3.1\(\times\) average latency improvement in collaborative mobile-cloud DNN execution, and by MEC/AI-RAN work that brings application hosting and AI execution closer to the radio access path \cite{satyanarayanan2017edge,kang2017,etsi2018mec5g,nvidia2024airan}. The longer-term trajectory is an edge continuum in which wearables, smartphones, building servers, base stations, and cloud resources are orchestrated as one logical computer rather than operated as isolated devices.

\subsection{Privacy-Preserving Personalization}

Personalization is essential for proactive computing, but it must be achieved without excessive data exposure. 
Federated learning, on-device adaptation, differential privacy, and secure aggregation can support user-specific models while reducing centralized data collection \cite{mcmahan2017communication,dwork2006calibrating,bonawitz2017practical}.

Future work should investigate how to personalize not only prediction models but also action policies, autonomy levels, and explanation styles.

\subsection{Foundation-Model-Based Proactive Agents}

Foundation models may enable proactive systems to reason over multimodal context, user history, and long-term goals. 
They can also generate natural-language explanations and support flexible interaction.

However, deploying foundation models in proactive settings raises challenges involving latency, hallucination, privacy, controllability, and verification. These issues are especially important because foundation models can generalize across tasks but also amplify reliability and governance risks \cite{bommasani2021foundation,bai2024hallucination,nist2023airmf}. 
Future systems should combine foundation-model reasoning with grounded sensor data, uncertainty estimation, constrained action policies, and adaptive sensing. The adaptive-sensing view is especially important: the Lens results suggest that changing how the world is sensed can improve accuracy dramatically even without modifying the downstream model, so proactive agents should learn when to increase sampling, switch modalities, change camera parameters, request on-premise computation, or offload to base-station/cloud resources \cite{baek2025adaptive,baek2025adaptive_sensing_position}.

\subsection{Human-in-the-Loop Autonomy}

Fully autonomous proactive action is not always desirable. 
Many applications require adjustable autonomy, where the system can shift between notification, recommendation, confirmation, and automation depending on risk and user preference.

Future research should explore human-in-the-loop mechanisms that preserve user agency while reducing interaction burden. 
This includes feedback learning, consent management, interactive correction, and transparent control interfaces.

\subsection{Sustainable and Event-Driven Proactivity}

Proactive computing should be designed for long-term sustainability. 
Rather than continuously sensing and computing, future systems should use event-driven, adaptive, and selective mechanisms. 
The system should allocate sensing and computation based on expected value, uncertainty, and resource state.

This direction connects proactive computing with efficient AI, tinyML, adaptive sensing, and energy-aware systems. It should also follow Green AI principles by reporting computational cost alongside task performance and by using compression or selective computation when possible \cite{han2015deep,schwartz2020green}.

\subsection{Benchmarks and Evaluation Protocols}

The field lacks standardized benchmarks for proactive computing. 
Future work should develop datasets, simulation environments, and user study protocols that evaluate not only prediction accuracy but also action quality, user acceptance, trust, privacy, physical correctness, temporal robustness, and sustainability. These protocols should incorporate technology acceptance constructs, trust-in-automation measurements, and AI risk management criteria rather than relying only on conventional machine-learning metrics \cite{davis1989perceived,venkatesh2003user,kohn2021measurement,nist2023airmf}.

Possible evaluation dimensions include:
\begin{itemize}
    \item prediction accuracy;
    \item intervention usefulness;
    \item timing quality;
    \item false intervention rate;
    \item user trust;
    \item controllability;
    \item energy consumption;
    \item privacy exposure.
\end{itemize}

Such evaluation frameworks are necessary for comparing proactive systems across domains. They should explicitly connect HAR and time-series benchmarks with physical-world benchmarks such as PhysBench, because proactive computing often fails at the boundary between temporal prediction and physical consequence: a model must not only infer what the user is doing, but also understand what will happen if the system acts \cite{yuan2024selfsupervisedhar,zhang2024sslts,chow2025physbench}.

\section{Conclusion}
\label{sec:conclusion}

Proactive computing is emerging from the convergence of pervasive sensing, AI-based understanding, distributed infrastructure, and physical actuation. It extends beyond reactive, context-aware, and predictive computing by enabling systems to initiate information delivery or actions before explicit user requests.

However, proactive computing is not merely a technical problem. Its success depends on the ability to integrate accurate sensing, reliable prediction, appropriate action, user acceptance, trust, privacy, accountability, and sustainability. The central challenge is not only to predict future user needs, but to decide when, how, and whether to act on behalf of users.

This survey framed proactive computing as a socio-technical system-level paradigm. We discussed its conceptual evolution, technological enablers, design space, technical challenges, socio-technical issues, and future directions. As computing systems continue to move closer to human life, proactive computing offers a promising but challenging path toward more personalized, timely, and intelligent services.

\bibliographystyle{plain} % 또는 unsrt, alpha 등 원하는 스타일 선택
\bibliography{references, eis-lab}

\end{document}